\documentclass[aps,prl,twocolumn,superscriptaddress]{revtex4-1}

\usepackage[pdftex]{graphicx}

\begin{document}

\title{Doping - dependent superconducting gap anisotropy in the two-dimensional pnictide Ca$_{10}$(Pt$_3$As$_8$)[(Fe$_{1-x}$Pt$_{x}$)$_2$As$_2$]$_5$}

\author{K.~Cho}
\affiliation{The Ames Laboratory, Ames, IA 50011, USA}

\author{M.~A.~Tanatar}
\affiliation{The Ames Laboratory, Ames, IA 50011, USA}

\author{H.~Kim}
\affiliation{The Ames Laboratory, Ames, IA 50011, USA}
\affiliation{Department of Physics and Astronomy, Iowa State University, Ames, IA 50011, USA}

\author{W.~E.~Straszheim}
\affiliation{The Ames Laboratory, Ames, IA 50011, USA}

\author{N.~Ni}
\affiliation{Department of Chemistry, Princeton University, Princeton, NJ 08544, USA}

\author{R.~J.~Cava}
\affiliation{Department of Chemistry, Princeton University, Princeton, NJ 08544, USA}

\author{R.~Prozorov}
\email[Corresponding author: ]{prozorov@ameslab.gov}
\affiliation{The Ames Laboratory, Ames, IA 50011, USA}
\affiliation{Department of Physics and Astronomy, Iowa State University, Ames, IA 50011, USA}

\date{3 November 2011}

\begin{abstract}
The characteristic features of Ca$_{10}$(Pt$_3$As$_8$)[(Fe$_{1-x}$Pt$_x$)$_2$As$_2$]$_5$ (``10-3-8'') superconductor are relatively high anisotropy and a clear separation of superconductivity and structural/magnetic transitions, which allows studying the superconducting gap without complications due to the coexisting order parameters. The London penetration depth, measured in underdoped single crystals of 10-3-8 ($x =$ 0.028, 0.041, 0.042, and 0.097), shows behavior remarkably similar to other Fe-based superconductors, exhibiting robust power-law, $\Delta \lambda(T) = A T^n$. The exponent $n$ decreases from 2.36 ($x =$ 0.097, close to optimal doping) to 1.7 ($x =$ 0.028, a heavily underdoped composition), suggesting that the superconducting gap becomes more anisotropic at the dome edge. A similar trend is found in low-anisotropy superconductors based on BaFe$_2$As$_2$ (``122"), implying that it is an intrinsic property of superconductivity in iron pnictides, unrelated to the coexistence of magnetic order and superconductivity or the anisotropy of the normal state. Overall this doping dependence is consistent with $s_{\pm}$ pairing competing with intra-band repulsion.
\end{abstract}



\maketitle

Since the discovery of the La(O$_{1-x}$F$_x$)FeAs (``1111") superconductor \cite{Kamihara2008JACS}, $T_c$ as high as 55 K have been reported in Sm[O$_{1-x}$F${_x}$]FeAs \cite{Ren2008CPL} triggering intense research. Subsequent studies of Fe-based superconductors (FeSC) indicated unconventional pairing and the coexistence of superconductivity with magnetism \cite{Ishida2009JPSJ,Paglione2010NP,Canfield2010ARCMP,Basov2011NP}. Similar to high-$T_c$ cuprates, FeSCs have layered chemical structures, with layers of Fe tetrahedrally coordinated by As or chalcogen anions (Se or Te) and determining the electronic structure at the Fermi level. Fe-As layers are alternatively stacked with alkali, alkaline earth or rare earth oxygen spacer layers. The low dimensionality of the electronic structure of the cuprates is believed to be responsible for their high $T_c$ and highly anisotropic gap (d-wave) \cite{2Dcuprates}. However, despite obviously layered structures, the electronic anisotropy of most - studied 122 compounds is rather low, with $\gamma_{H} \equiv H_{c2,ab}/H_{c2,c} \sim 2-3$ at  $T = T_c$ and decreasing upon cooling \cite{Gurevich2011}. The 122 pnictides also exhibit clear evolution of the superconducting gap with doping from isotropic at the optimal concentrations towards nodal structure at the dome edges \cite{ProzorovKoganROPP2011}. To check whether electronic anisotropy plays a role in the structure of the superconducting gap, more anisotropic materials with controlled doping are needed. Unfortunately, in higher anisotropy 1111 system, with $\gamma_H(T_c) \approx 7$, control of the doping level in single crystals in not achieved yet.

\begin{figure}[tb]
\includegraphics[width=7.2cm]{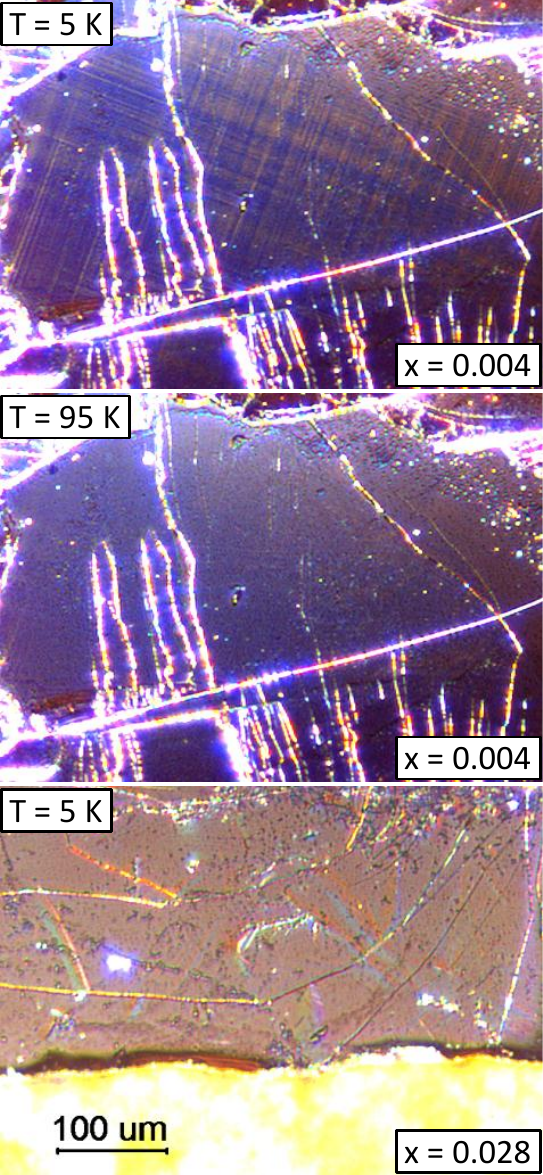}
\caption{\label{domains} (Color online) Polarized-light images of single crystals of the 10-3-8 phase with $x=0.004$ at 5 K showing a clear twin-domain pattern (top), and above the transition, at 95 K, with no domains (middle). The most underdoped superconducting composition, $x=0.028$, does not show domains down to 5 K (bottom).
}
\end{figure}

The intermediary spacer has been one of the key parameters in engineering high-$T_c$ superconductors. In cuprates, the highest $T_c$s in (Bi, Tl, Hg) - based superconductors were facilitated by enhancing the CuO$_2$ plane coupling \cite{Subramanian1988Science, Subramanian1988Nature, Torardi1988Science, Tokiwa-Yamamoto1993PC}. In Fe-based superconductors,  various spacer layers are also used to control the interlayer coupling and the crystal structure is categorized based on the intermediary spacer layers, such as $A$FeAs ($A$ - alkali ion), $AE$Fe$_2$As$_2$ ($AE$ - alkali-earth ion), $RE$FeAsO ($RE$ - rare-earth ion) and $Ox$Fe$_2$As$_2$ ($Ox$ - complex metal oxide) \cite{Zhu2009PRB_a, Zhu2009PRB_b, Kawaguchi2010APE, Ogino2010SST, Shirage2010APL}.

Recently a new family of FeSCs with PtAs intermediary spacer layers has been reported in a Ca-Fe-Pt-As system \cite{Nohara2011workshop, Kudo2011workshop} and now high purity single crystals for the superconductors Ca$_{10}$(Pt$_n$As$_8$)[(Fe$_{1-x}$Pt$_x$)$_2$As$_2$]$_5$ where n = 3 (the ``10-3-8'' phase) and 4 (the ``10-4-8" phase) are available \cite{Ni2011arXiv, Kakiya2011arXiv, Lohnert2011arXiv}. The 10-3-8 phase with triclinic symmetry (which is rare in superconductors) shows superconducting $T_c$ up to 13 K upon Pt-doping. The superconductivity of the 10-4-8 phase, which has a tetragonal symmetry, stabilizes at a higher $T_c$ of 38 K \cite{Kakiya2011arXiv}. The anisotropy of 10-3-8 phase is even higher than in 1111 compounds, $\gamma_H(T_c) \sim 10$ \cite{Ni2011arXiv}, but contrary to 1111, there is a good control over the doping level. An interesting feature of the 10-3-8 phase, suggested by the transport measurements \cite{Ni2011arXiv} and supported by our direct imaging of structural domains shown in Fig.~\ref{domains}, is a clear separation of structural instability and superconductivity. This is distinctly different from the 122 pnictides, where these two order parameters coexist up to the optimal doping \cite{Prozorov2009intrinsic}. Therefore, 10-3-8 system provides a new opportunity to study the evolution of the gap structure in different parts of the superconducting dome without the possible influence of competing magnetic and structural orders.

In this work, we studied 10-3-8 crystals in the underdoped regime up to optimal doping. The low-temperature penetration depth exhibits power-law variation, $\Delta \lambda = A T^{n}$, with the exponent $n$ decreasing towards the edge of the dome. This behavior is similar to low-anisotropy BaK122 (hole doped) \cite{Kim2011BaKLambda} and BaCo122 (electron doped)\cite{Gordon2010Lambda0}. We conclude that neither the normal-state anisotropy nor the coexistence of superconductivity and magnetism play a significant role in determining the evolution of the superconducting gap structure in FeSCs.

Single crystals of Ca$_{10}$(Pt$_3$As$_8$)((Fe$_{1-x}$Pt$_x$)$_2$As$_2$)$_5$ were synthesized as described elsewhere \cite{ Ni2011arXiv}. The compositions of six samples were determined with wavelength dispersive spectroscopy (WDS) electron probe microanalysis as $x=$ 0.004$\pm$0.002, 0.018$\pm$0.002, 0.028$\pm$0.003, 0.041$\pm$0.002, 0.042$\pm$0.002, and 0.097$\pm$0.002. The in-plane London penetration depth, $\lambda (T)$, was measured using a self-oscillating tunnel-diode resonator (TDR) technique \cite{Prozorov2006SST, Van-Degrift1975RSI, Prozorov2000PRB}. The sample was placed with its c-axis along the direction of ac-field, $H_{ac}$, induced by the inductor coil. Since $H_{ac} \sim$ 20 mOe is much weaker than the first critical field, the sample is in the Meissner state, so its magnetic response is determined by the London penetration depth, $\lambda(T)$. Details of the measurements and calibration can be found elsewhere \cite{Prozorov2006SST}. Polarized light imaging of the structural domains, shown in Fig.~\ref{domains}, was done with a Leica microscope in a flow-type optical $^4$He cryostat, for details see \cite{Tanatar2009domains,Prozorov2009intrinsic}.

Figure~\ref{domains} shows polarized-light optical images for the undoped,  $x=0.004$ (top panel - at 5 K, and middle panel - at 95 K), and most underdoped, $x=0.028$ (at 5 K - lower panel) compositions. The mesh-like contrast clearly visible in the top panel is due to the formation of structural domains. The polarization plane rotates differently upon reflection from the neighboring twins, thus resulting in a difference in the light contrast as observed through an analyzer.  Usually, such domains are formed upon lowering the symmetry of the lattice. In most pnictides this is a transition from the tetragonal to the orthorhombic phase and is also accompanied by an antiferromagnetic transition \cite{Tanatar2009domains,Prozorov2009intrinsic}. In the present case, we already start with the lowest symmetry triclinic system and the only possibility to form domains is to have another transition that would cause additional stress. More information on structural domains in triclinic systems can be found elsewhere \cite{triclinic}. Long-range antiferromagnetic ordering would cause such stresses, via magnetostrictive coupling, which is very pronounced in the pnictides. In the undoped composition, the present observation of domains coincides with the feature observed in the resistivity on samples from the same batch \cite{Ni2011arXiv}. Note that despite its triclinic crystal structure, a four - fold symmetry of the electronic structure was found in angle resolved photoemission measurements \cite{Neupane2011arXiv}, suggesting that triclinic distortion plays minor role. The lower panel in Fig.~\ref{domains} shows the most underdoped 10-3-8 composition with  $x=0.028$ imaged at 5 K. We do not observe any domains and the picture does not change at room temperature. Coupled with the absence of any features in the resistivity \cite{Ni2011arXiv}, we conclude that the 10-3-8 system does not exhibit the coexistence of superconductivity and structural or magnetic instability.

\begin{figure}[tb]
\includegraphics[width=8.5cm]{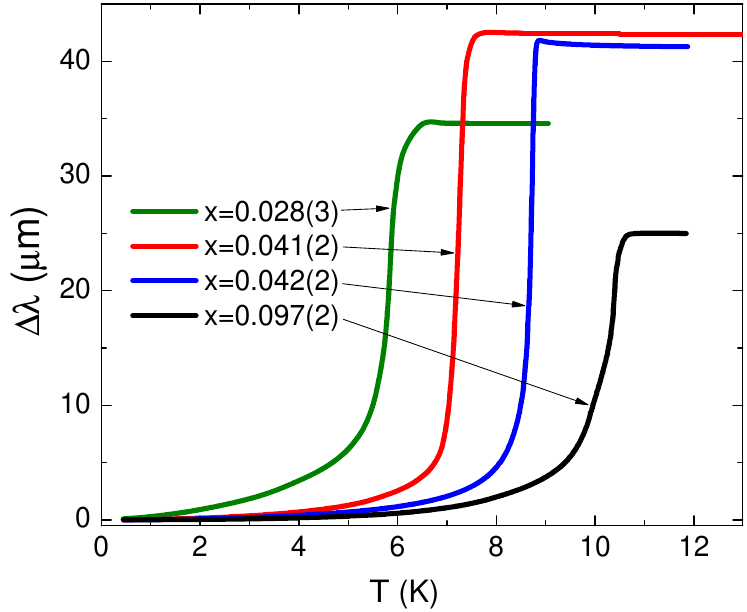}
\caption{\label{fullrange} (Color online) Variation of the London penetration depth, $\Delta \lambda(T)$, in the full temperature range for four underdoped compositions of the 10-3-8 phase. There is clear monotonic increase of $T_c$ with Pt content indicated in the legend.
}
\end{figure}

Figure~\ref{fullrange} shows the variation of the London penetration depth, $\Delta \lambda(T)$, during a temperature sweep through the superconducting transition in 10-3-8 single crystals with $x =$ 0.028, 0.041, 0.042, and 0.097. All samples show clear sharp superconducting transitions, with $T_c$ monotonically increasing with $x$, consistent with the transport measurements of the crystals from the same batches \cite{Ni2011arXiv}.

\begin{figure}[tb]
\includegraphics[width=8.5cm]{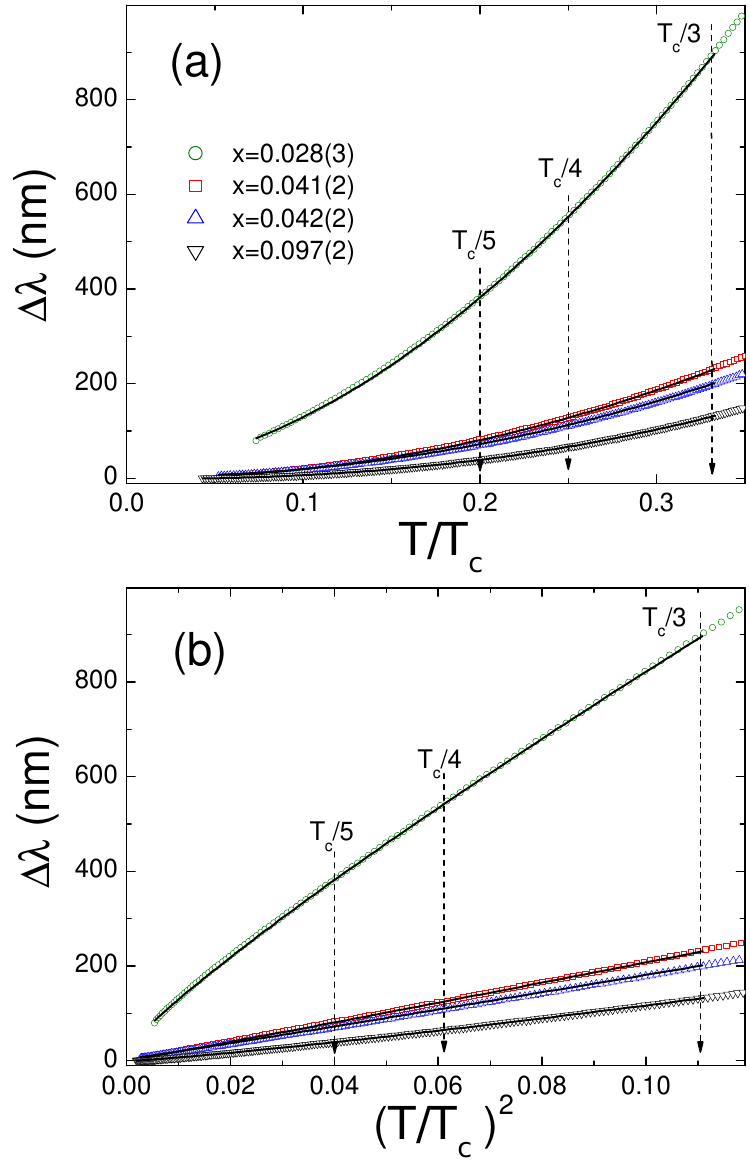}
\caption{\label{range} (Color online) Low temperature variation of the London penetration depth, $\Delta \lambda$, plotted against (a) $T/T_c$ and (b) $(T/T_c)^2$ for Pt-doping $x =$ 0.028, 0.041, 0.042, and 0.097. The vertical dashed lines indicate the upper limits of the fitting ranges, $T_c/5$, $T_c/4$ and $T_c/3$. The solid lines are representative fits to $\Delta \lambda = A T^n$ for each doping, conducted with the upper limit of $T_c/3$. The resulting exponents $n$ for all three fitting ranges are shown in Fig.~\ref{PD} (a).}
\end{figure}

Figure~\ref{range} shows the low-temperature variation of $\Delta \lambda$ plotted against (a) linear, $T/T_c$, and (b) quadratic, $(T/T_c)^2$, normalized temperature scales. For the quantitative analysis, $\Delta \lambda (T)$ was fitted to a power-law equation, $\Delta \lambda (T) = A T^n$. To examine the robustness of the fits, for each doping level multiple fittings were performed for three different temperature ranges with a fixed lower limit of 0.5 K and a variable upper limit of $T_c$/3, $T_c$/4, and $T_c$/5 (indicated by dashed lines in Fig.~\ref{range}). The symbols in Fig.~\ref{range} are the data and the solid lines show representative fits for different doping levels for the upper temperature limit of $T_c$/3.

\begin{figure}[tb]
\includegraphics[width=8.5cm]{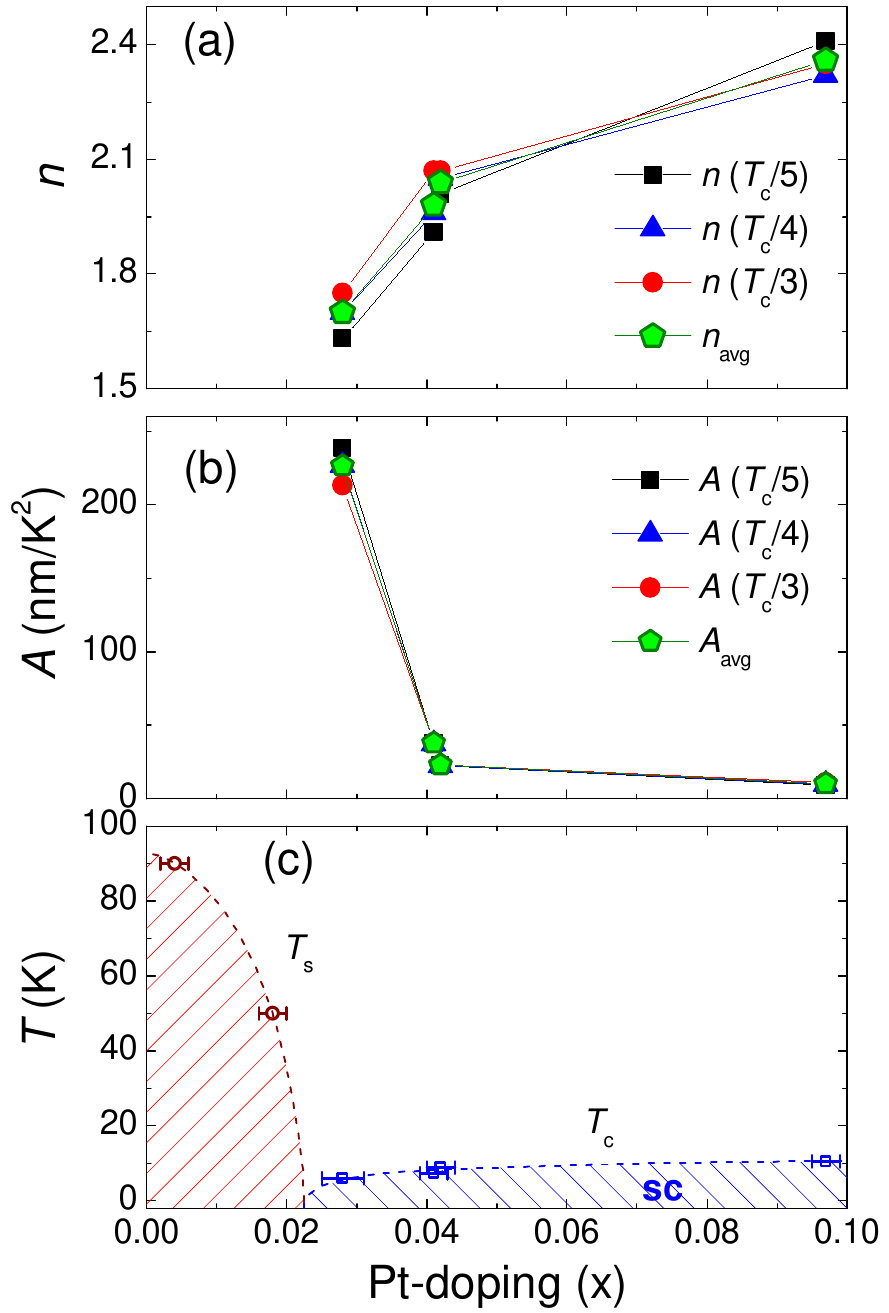}
\caption{\label{PD} (Color online) Results of the power-law fits with three different upper limits, indicated by dashed lines in Fig.~\ref{range} are shown along with the average values. Panel (a): the exponent $n$, obtained by keeping $A$ and $n$ as free parameters. Panel (b): the pre-factor $A$, obtained at a fixed $n=2$. Panel (c) shows the doping phase diagram with the magnetic (M) and superconducting (SC) phases clearly separated as a function of excess Pt content $x$. $T_s$ measured from resistivity \cite{Ni2011arXiv} and $T_c$ from TDR (this work).}
\end{figure}

The resulting exponents $n$ for all three fitting ranges are shown in Fig.~\ref{PD} (a). Figure~\ref{PD} (b) shows the prefactor $A$ obtained at a fixed $n=2$ for different fitting ranges. To compare the behavior of the London penetration depth in samples of different doping levels, the average values (obtained for three different fitting ranges), $n_{avg}$ and $A_{avg}$ will be discussed. As shown in Fig.~\ref{PD} (a), $n_{avg}$ increases from 1.7 to 2.36 as the doping level increases from 0.028 to 0.097. Since the maximum of $T_c$ is expected to occur near or above $x = 0.097$, it can be concluded that the exponent, $n$, increases towards optimal doping. At the same time, the prefactor, $A$, decreases with the increase of the doping level. Note that $A_{avg}$ shows a dramatic five-fold decrease from the lowest doping at $x =$ 0.028 to  its value at the optimal $x =$ 0.097. This behavior is naturally explained by a larger gap anisotropy in the more underdoped compositions, which leads to a larger density of low-energy quasiparticles thermally excited over the gap minima, causing $A$ to increase dramatically and the exponent $n$ to decrease. The ultimate limit will be formation of nodes with $n=1$ in the clean limit.

A similar doping-dependent evolution of the power-law London penetration depth was found in BaCo122 \cite{Gordon2010Lambda0}. For that compound, it was suggested that the underdoped side is significantly affected by co-existing magnetic order and was naturally explained by an increasing gap anisotropy when moving towards the edge of the ``superconducting dome'', consistent with thermal conductivity \cite{Tanatar2010TC,Reid2010c-axis} and specific heat \cite{Budko2009Cp} results. In the present case of the 10-3-8 system where magnetism and superconductivity are separated in the phase diagram, shown in panel (c) of Fig.~\ref{PD}, this doping-dependent evolution of $n$ and $A$ suggests that the development of the anisotropic gap structure upon departure from optimal doping is a universal intrinsic feature of iron-pnictides, and is not directly related to the structural and electronic anisotropies or to the coexistence of magnetism and superconductivity.

One possible explanation for the variation in the superconductivity parameters is that the pairing mechanism is different in different parts of the phase diagram, for example evolving from magnetic- to orbital-fluctuation- mediated superconductivity. Recent theoretical treatments that take into account  a variety of experimental results for many different systems, however, suggest that the physics is universal and can be understood within the same universal pairing scenario based on competing inter-band coupling and intra-band Coulomb repulsion and pair-breaking impurity scattering \cite{Chubukov2011review,Hirschfeld2011ROPP}.

In conclusion, the London penetration depth, $\lambda (T)$, was measured in single crystals of Ca$_{10}$(Pt$_3$As$_8$)[(Fe$_{1-x}$Pt$_x$)$_2$As$_2$]$_5$ (10-3-8) with different levels of excess Pt content, $x$. The superconducting transition temperature, $T_c$, increases monotonically reaching 10.5 K for $x=0.097$. The power-law fit to the low temperature part of $\Delta \lambda$(T) shows that the average exponent, $n_{avg}$, varies from 1.7 to 2.36, which can be naturally explained by an increasing anisotropy of the superconducting gap when moving towards the edges of the superconducting dome. This behavior is not a consequence of the coexistence of superconductivity and magnetism and chemical or electronic anisotropy. It is a universal and robust property of iron pnictide superconductors and, most likely, comes from the multiband physics of the superconducting pairing.

\begin{acknowledgments}
We thank Andrey Chubukov and Peter Hirschfeld for useful discussions. Work at the Ames Laboratory was supported by the Department of Energy-Basic Energy Sciences under Contract No. DE-AC02-07CH11358. The work at Princeton University was supported by the AFOSR MURI on superconductivity.
\end{acknowledgments}

\end{document}